\documentclass[aps,prd, nofootinbib, notitlepage,onecolumn]{revtex4-1}
\usepackage{amsmath,amsfonts,amssymb}
\usepackage{bm}
\usepackage{framed}
\usepackage{latexsym}
\usepackage{graphicx,epsfig}
\usepackage{psfrag}
\usepackage{amsthm}
\usepackage{color}
\usepackage{braket}
\usepackage{indent first}
\usepackage[normalem]{ulem}
\usepackage{fancyhdr}
\usepackage{hyperref}
\usepackage[titletoc,title]{appendix}
 \usepackage{environ}
\usepackage{mathtools}
\usepackage{float}
\usepackage[makeroom]{cancel}
\newcommand{\be}{\begin{equation}}
\newcommand{\ee}{\end{equation}}
\newcommand{\bea}{\begin{eqnarray}}
\newcommand{\eea}{\end{eqnarray}}

\NewEnviron{eqn}{
\begin{align}
\begin{split}
  \BODY
\end{split}
\end{align}
}

\NewEnviron{eqn*}{
\begin{align*}
\begin{split}
  \BODY
\end{split}
\end{align*}
}

\begin{document}
\title{Dark Matter as an effect of a minimal length}

\author{Pasquale Bosso} 
\email{bosso.pasquale@gmail.com}
\affiliation{Theoretical Physics Group and Quantum Alberta, Department of Physics and Astronomy,
University of Lethbridge,
4401 University Drive, Lethbridge,
Alberta, T1K 3M4, Canada}

\author{Mitja Fridman}
\email{fridmanm@uleth.ca}
\affiliation{Theoretical Physics Group and Quantum Alberta, Department of Physics and Astronomy,
University of Lethbridge,
4401 University Drive, Lethbridge,
Alberta, T1K 3M4, Canada}

\author{Giuseppe Gaetano Luciano} \email{gluciano@sa.infn.it}
\affiliation{Dipartimento di Fisica E.R. Caianiello, Universit\`a di Salerno, Via Giovanni Paolo II, 132 - 84084 Fisciano, Salerno, Italy} 
\affiliation{INFN, Sezione di Napoli, Gruppo collegato di Salerno, 84084 Fisciano, Salerno, Italy}

\begin{abstract}
In this work, we consider the implications of a phenomenological model of quantum gravitational effects related to a minimal length, implemented via the Generalized Uncertainty Principle.
Such effects are applied to
the Bekenstein-Hawking entropy to derive a modified law of gravity through Verlinde's conjecture.
Implications on galactic scales, and in particular on the shape of rotational curves, are investigated, exploring the possibility to mimic dark matter-like effects via a minimal length.
\end{abstract}
\maketitle
%

\section{Introduction}

Dark Matter (DM) is one of the dominant components in the energy budget of the universe. Evidence for its existence ranges from galaxy clusters, rotational curves of galaxies, gravitational lensing, all to the Cosmic Microwave Background (CMB) \cite{Freese:2008cz,Drees:2012ji,Arbey:2021gdg}.
However, effects related to DM have not been observed on the scale of the Solar System, whereas they become significant on galactic and intergalactic scales.
Nonetheless, the nature of DM remains one of the most debated problems in Physics up to this day.
Several proposals and speculations concerning DM have been put forward, among which are MACHOs, WIMPs, Axions, Sterile neutrinos and Modified Newtonian Dynamics (MOND) \cite{Freese:2008cz,Drees:2012ji,Arbey:2021gdg,Oks:2021hef}.
In this work, we propose an alternative explanation for DM.
Specifically, we argue that the phenomenology related to DM can be partially described in terms of quantum gravitational effects.

The development of a theory of Quantum Gravity (QG) is an open problem in Physics.
Several candidate theories have been proposed and numerous thought experiments have shaped the expected features of such a theory.
However, none could have been directly tested due to current experimental and technological limitations.
For this reason, phenomenological approaches have become some of the main tools to tackle the problem of QG \cite{CM1,CM2,ADV0,CM3,CM4,CM5,CM6,CM7} (see~\cite{Addazi} for a recent review).
Such approaches usually consider the implications of features of a full QG theory on lower energy scales, possibly accessible to current experiments and observations.
Among such features, a common one is the existence of a fundamental minimal length.
Such a minimal length strongly opposes the traditional Heisenberg Uncertainty Principle of quantum mechanics, which should be properly  modified approaching the QG scale.
The set of models corresponding to a modified uncertainty relation is collectively referred to as Generalized Uncertainty Principle (GUP) \cite{ADV0,GUP1,GUP2,GUP3,Kempf:1996fz,GUP4,KMM,FC,IPK,SLV,Bosso:2016ycv,KP,Bosso:2018uus,Gnatenko:2019flg,Giardino:2020myz,Bosso:2020aqm,Das:2021skl,BossoLuc,Luciano:2021ndh,AHG}.
Such models are inspired from candidate theories of quantum gravity, such as string theory and loop quantum gravity, in which an effective minimal observable length appears in scattering experiments or as a structural feature of space time.
The phenomenological implications are then accounted for in terms of a minimal uncertainty in position or non-commutativity of space-time.
The commutator corresponding to one of most common GUP models can be cast as \cite{KMM,Bosso:2020aqm}
\begin{equation}
 \label{gup}
     [x_i,p_j]=i\hbar\left[\delta_{ij}+\beta\left(p^2\delta_{ij}+2p_ip_j\right)\right]~,
 \end{equation}
where $x_i$ and $p_j$ are the position and momentum operators, respectively, $\beta \equiv \beta_0/(M_{P}c)^2$, where $\beta_0$ is a dimensionless parameter and $M_{P}=\sqrt{\hbar c/G}$ is the Planck mass.
Such a model implies a modification of the uncertainty relation between position and momentum, as found using the Schr\"odinger-Robertson relation, and thus to a minimal uncertainty in position.
For example, in the one-dimensional case, the modified uncertainty relation reads \cite{KMM}
\begin{eqnarray}
\Delta x \Delta p \gtrsim \hbar\left[1+3\beta\Delta p^2\right].
\end{eqnarray}

In this paper, we propose how minimal length phenomenology can give rise to features similar to DM on galactic scales.
Specifically, we deduce effects from GUP that contribute to the flatness of rotational curves.
Such effects are obtained as a consequence of the modifications to the Bekenstein Hawking entropy through the holographic principle, induced by GUP \cite{Cai:2008ys,Zhu:2008cg,AA,Giardino:2020myz,Das:2021nbq,ScardBek}.
Therefore, we obtain corrections to the corresponding entropic force due to the presence of a minimal length.
Based on Verlinde's conjecture~\cite{Verlinde:2010hp}, such a modified entropic force turns out as a modified Newton's law of gravity, thus providing a basis to study the implications of a minimal length on gravitational systems.
Specifically, we require the holographic principle to hold.
That is, we consider spherically symmetric surfaces of area $A = 4 \pi R^2$ separating points in space.
Such surfaces behave as the natural place to store information about particles that are inside the surfaces and that can move from one side to another.
In this way, the information about the location of particles is stored in discrete bits on the surfaces.
This naturally leads to assume that the number $N_b$ of bits on the screens can be approximated with the number of particles $N_p$ enclosed by the surfaces, i.e. $N_b\sim N_p=N$.
Then, the total number $N$ of bits of the system, which is measured by its entropy $S$, can be naturally assumed to be proportional to the surface area $A$, i.e. $N\sim S\sim A$.
The total energy $E$ of the system inside the surfaces is distributed on such bits and is related to the surface temperature by the equipartition theorem (or the GUP-modified version thereof).
Such energy can be written in terms of the uniformly distributed mass $M$ inside the surface as $E=Mc^2$. 
Notice that the above reasoning can in principle be applied to any mass distribution, as long as one defines a proper holographic screen of radius $R$ containing the whole distribution.

It turns out that a distance dependence for the GUP parameter $\beta_0$ must be assumed to provide a reasonable mechanism to study minimal length effects on rotational curves of galaxies.
Such dependence has been proposed in other works as well (see, e.g., Ref. \cite{Ong:2018zqn}) and suggested by the different estimations of the GUP parameter in tabletop experiments, where $\beta_0>0$ \cite{IPK,SLV,Bosso:2016ycv,KP,Das:2021skl}, and astrophysical/cosmological observations, where $\beta_0<0$  \cite{Das:2021nbq,Ong:2018zqn,Nenmeli:2021orl,JizbaScard,BuoninfCorp,JizbaLamb} (see also~\cite{RevLuc} for a recent review). 

The paper is structured as follows:
in section \ref{ET}, a modification to the equipartition theorem due to GUP is introduced;
in section \ref{DM}, a modified Newton's law of gravity is derived from the GUP-modified equipartition theorem and the Bekenstein-Hawking entropy;
in section \ref{conc}, we summarize our results and include some final remarks.

\section{GUP-modified equipartition theorem}
\label{ET}

One of the reasons to introduce DM is the flatness of galactic rotational curves, which deviate from the behavior predicted based on Kepler's model considering only luminous matter.
In particular, Kepler's laws predict that the orbital velocities for stars outside the bulge decreases as the square root of the distance from the center, $v(R) \propto 1/\sqrt{R}$.
The observation that the orbital velocities are approximately independent of the distance from the center, $v(R)\propto const.$, even at distances comparable with the galactic radius and beyond, suggests that either Newton's law of gravity does not hold at such scales, or that non-visible matter, present in galaxies, affects stellar dynamics.
As mentioned in the Introduction, here we will consider the former, with the intent of studying the implications of a minimal length on galactic rotational curves.
Specifically, following~\cite{Verlinde:2010hp}, we introduce the gravitational force $F$ as an entropic force
\begin{equation}
    \label{entforce}
    F\Delta x=T\Delta S~,
\end{equation}
where $\Delta x=\eta\lambda_C=\eta\frac{\hbar}{mc}$ is the displacement from the source of a test particle in terms of its Compton wavelength $\lambda_C$, $m$ is the mass of the test particle, $\eta$ is the modification factor ($\eta=1$ if no modification), $T$ is the effective temperature at a given radius, and $\Delta S=2\pi k_B$ is the minimal change in entropy, as stated by information theory \cite{Adami:2004mx}, with $k_B$ being the Boltzmann constant.
The effective temperature $T$ can be expressed in terms of the energy $E$ of a thermodynamical system via the equipartition theorem.
In our case, we assume that the mass of the system is contained within a sphere of radius $R$.
However, when a minimal length is introduced considering the effects of GUP on statistical mechanics, the equipartition theorem presents some corrections.
To see this, we first notice that the GUP-modified density of states in three spatial dimension reads as \cite{Das:2021skl}
\begin{equation}
    \label{mdens}
    g(\varepsilon)=\frac{V(2m)^{3\slash2}\varepsilon^{1\slash2}}{4\pi^2\hbar^3}\left(1-\frac{75}{4}\beta m\varepsilon\right)~,
\end{equation}
where $V$ is the volume of the system and $\varepsilon$ is the single particle energy.
Since the results in \cite{Das:2021skl} are valid up to order $\beta$, the density of states presented above, as well as its consequences, are understood to hold up to the same order.
We anticipate that the volume $V$ will cancel out in our considerations and does not affect the results.
Furthermore, it is worth mentioning that GUP is not expected to modify the value of geometrical quantities such as volumes or areas.
Since GUP is a phenomenological model of quantum mechanics including a minimal uncertainty in position, GUP only affects the precision with which particles are localized, and therefore the precision with which geometrical quantities are determined, not their actual value.
Returning to the expression above, it is worth noticing that it reduces to the standard density of states for $\beta \to 0$ and that such a result is quantum in nature since it is based
on the quantum energy spectrum of a particle in a box. In the classical limit $\varepsilon-\mu\gg T$, where $\mu$ is the chemical potential, and assuming no particles are added or removed from the system, $\mu=0$, the Bose-Einstein and Fermi-Dirac distributions reduce to
\begin{equation}
    \label{classdist}
    f_{_\mathrm{BE,FD}}(\varepsilon) = \frac{1}{\exp{\left(\frac{\varepsilon-\mu}{k_BT}\right)}\mp1} \approx f(\varepsilon)=\exp{\left(-\frac{\varepsilon}{k_BT}\right)}~,
\end{equation}
where the $-$ and the $+$ signs refer to the Bose-Einstein and Fermi-Dirac distributions, respectively.
The limit on the right hand side of Eq. (\ref{classdist}) is the Maxwell-Boltzmann distribution.

To proceed further, we compute the number of particles in the system by considering the following ensemble average using the GUP-modified density of states from Eq. \eqref{mdens} and the classical limit for the particle distribution in Eq. \eqref{classdist}.
We then obtain
\begin{equation}
    \label{statnum}
    N = \int_0^\infty g(\varepsilon)\,f_{_\mathrm{BE,FD}}(\varepsilon)\,\mathrm{d}\varepsilon
    \simeq \frac{V(2m)^{3/2}}{8\pi^{3/2}\hbar^3}(k_BT)^{3/2} \left[1-\frac{225}{8}\beta m(k_BT)\right]~,
\end{equation}
where the additional term with $\beta$ represents the GUP correction to the number of particles in the system, given the temperature $T$ of the system and the mass of the constituent particles $m$.
The energy of the system is obtained in a similar manner as the number of particles in the system from Eq. (\ref{statnum}).
In this case, we find
\begin{equation}
    \label{staten}
    E = \int_0^\infty \varepsilon \, g(\varepsilon)\,f_{_\mathrm{BE,FD}}(\varepsilon)\,\mathrm{d}\varepsilon
    \simeq \frac{3V(2m)^{3/2}}{16\pi^{3/2}\hbar^3}(k_BT)^{5/2} \left[1-\frac{375}{8}\beta m(k_BT)\right]~.
\end{equation}
The above formula represents the thermal energy of a system in three spatial dimensions.
We can recast the expression for the thermal energy in a more familiar form by combining Eqs. (\ref{statnum}) and (\ref{staten}) to obtain the GUP-modified equipartition theorem, which reads as
\begin{eqnarray}
    E = \frac{n_s}{2}Nk_BT \left[1-\frac{75}{4}\beta m (k_BT)\right]~,
\end{eqnarray}
in $n_s$-spatial dimensions.
We are going to use this  expression with $n_s=1$, since the only relevant spatial degree of freedom in the system contributing to the entropic force is the radial one.
We can then find an expression of the temperature $T$ as a function of the energy $E$ up to first order in $\beta$, \emph{i.e.}
\begin{eqnarray}
    \label{temp}
    T \sim \frac{2E}{k_BN}+\beta\frac{75mE^2}{k_BN^2}~.
\end{eqnarray}
Since this expression is derived from a quadratic equation, two solutions are in principle allowed.
However, only the solution with the minus sign is considered since it is the only one which returns the standard case as $\beta \longrightarrow 0$.
For the case of the entropic force, the energy in Eq. (\ref{temp}) is not simply the thermal energy of particles in a given volume, but the total energy of the system in that volume.

\section{GUP-modified law of Gravity}
\label{DM}

As shown in Ref. \cite{Verlinde:2010hp}, one can derive Newton's law of gravity as an entropic force. The same procedure is applied here, with the difference that one includes GUP corrections everywhere they apply.
A similar consideration has been discussed in Ref. \cite{Sheykhi:2019bsh}, where a Tsallis entropy modification to the Bekenstein-Hawking (BH) entropy has been used to derive the modified law of gravity.
Such an entropy can be modified by GUP through the holographic principle as well \cite{AA,Das:2021nbq}.
It can then be written as
\begin{equation}
    \label{bhe}
    S=\frac{c^3 k_B}{8\hbar G}\left[A+\sqrt{A^2-\beta^*A}-\frac{\beta^*}{2}\ln{\left(1-\frac{2}{\beta^*}\left(A+\sqrt{A^2-\beta^*A}\right)\right)}\right]~,
\end{equation}
where $A$ is the area of the horizon and $\beta^*=12\pi\hbar^2\beta$.
Such a modification implies a deformation of the temperature of the system.
To see this, first we notice that the total number $N$ of bits of information on the surface of the holographic horizon can be expressed in terms of the horizon entropy as \cite{Verlinde:2010hp}
\begin{equation}
    \label{number}
    N=\frac{4S}{k_B}~.
\end{equation}
Thus, substituting Eq. \eqref{number} and Eq. \eqref{bhe} in Eq. \eqref{temp}, we get
\begin{equation}
    \label{guptemp}
    T=\frac{2\hbar GE}{k_Bc^3A}+\beta_0\left[\frac{6\pi\hbar^3GE}{k_BM_p^2c^5A^2}\left(1+\ln{\left(2-\frac{M_p^2c^2A}{3\pi\hbar^2\beta_0}\right)}\right)+\frac{75\hbar^2G^2mE^2}{k_BM_p^2c^8A^2}\right]~.
\end{equation}
It is worth noticing that the temperature $T$ depends on the area $A$ of the holographic horizon and the total energy $E$ of the system.
These quantities are related to the radius $R$ of the horizon and the mass $M$ contained within that radius through $A=4\pi R^2$ and $E=Mc^2$, respectively. 
The temperature from Eq. (\ref{guptemp}), expressed in terms of $R$ and $M$, and the aforementioned $\Delta x$ and $\Delta S$ are substituted in Eq. \eqref{entforce}, where the GUP modified Compton wavelength with $\eta=1+\beta_0\frac{m^2}{M_p^2}$ \cite{Carr:2020hiz} is considered.
Up to leading orders in $\beta_0$, we obtain the GUP-modified law of gravity
\begin{equation}
    \label{gravforce}
    F=\frac{GmM}{R^2}+\beta_0\left[\frac{3\hbar^2GmM}{4M_p^2c^2R^4}\left(1+\ln{\left(2-\frac{4M_p^2c^2R^2}{3\hbar^2\beta_0}\right)}\right)+\frac{75\hbar G^2 m^2M^2}{8M_p^2c^3R^4}-\frac{Gm^3M}{M_p^2R^2}\right]~.
\end{equation}
Using the expression for the centripetal force $F=m\frac{v^2}{R}$ at radius $R$, we can then find
the GUP-modified rotational velocity
\begin{equation}
    \label{rotvel}
    v=\sqrt{\frac{GM}{R}+\beta_0\left[\frac{3\hbar^2GM}{4M_p^2c^2R^3}\left(1+\ln{\left(2-\frac{4M_p^2c^2R^2}{3\hbar^2\beta_0}\right)}\right)+\frac{75\hbar G^2mM^2}{8M_p^2c^3R^3}-\frac{Gm^2M}{M_p^2R}\right]}~.
\end{equation}
The test particle, of mass $m$, can in principle be anything, from a subatomic particle to a large star.
However, when a composite system is considered, GUP effects are reduced by the number $n$ of constituents \cite{Amelino-Camelia:2013fxa}.
In the present case, this amounts to replacing the modification parameter $\beta_0$ by the reduced parameter scaling with the square of the inverse number of constituents, i.e. $\beta_0 \to \beta_0 / n^2$.
Note that $n$ corresponds to the number of constituent particles of the test mass which is different than $N$, introduced in Eq. (\ref{statnum}), which corresponds to the number of bits of information on the holographic screen.

The last term in Eq. (\ref{gravforce}) dominates at galactic distances, compared to other correction terms, as can be easily verified.
Therefore Eq.(\ref{gravforce}) can be rewritten as 
\begin{equation}
    F\simeq \frac{G m_{eff} M}{R^2}~,
\end{equation}
where we have defined an effective gravitational mass $m_{eff}=m\left[1-\frac{\beta_0}{n^2}\frac{m^2}{M_p^2}\right]$. We notice that this implies a potential GUP-induced violation of the weak equivalence principle, since $m_{eff}\equiv m_g\neq m\equiv m_i$, where $m_g$ is the gravitational mass and $m_i$ is the inertial mass (see also \cite{Casadio:2020rsj}).
Since in our framework $\beta_0<0$, we have $m_g>m_i$, which might partially justify the higher galaxy rotation velocities with respect to standard cosmological predictions.

From Eq.(\ref{rotvel}), the GUP-modified velocity for large distances is then
\begin{equation}
    v\simeq\sqrt{\frac{G M}{R}\left[1-\frac{\beta_0}{n^2}\frac{m^2}{M_p^2}\right]}\,\,\,\,\,\,\,\,\,\,\,\,\mathrm{with}\,\,\,\,\,\,\,\,\,\,\,\,\beta_0<0~.
    \label{eqn:orbital_velocity_GUP}
\end{equation}
We point out here that other terms in Eqs. (\ref{gravforce}) and (\ref{rotvel}) dominate at smaller scales, where the GUP effects are significantly smaller. 
Furthermore, it is worth noticing that, due to the scaling of the modification parameter in Eq. \eqref{eqn:orbital_velocity_GUP}, the mass of a star orbiting with velocity $v$ is scaled by the number of fundamental constituents.
Assuming that such constituents correspond to the elements in the plasma (mostly electrons and protons for a typical main sequence star), the quantity $m/n$ is of the order of the proton mass regardless of the actual values of $m$ and $n$.
Such assumption will be considered in the estimations below.

\begin{figure}[H]
    \centering
    \includegraphics{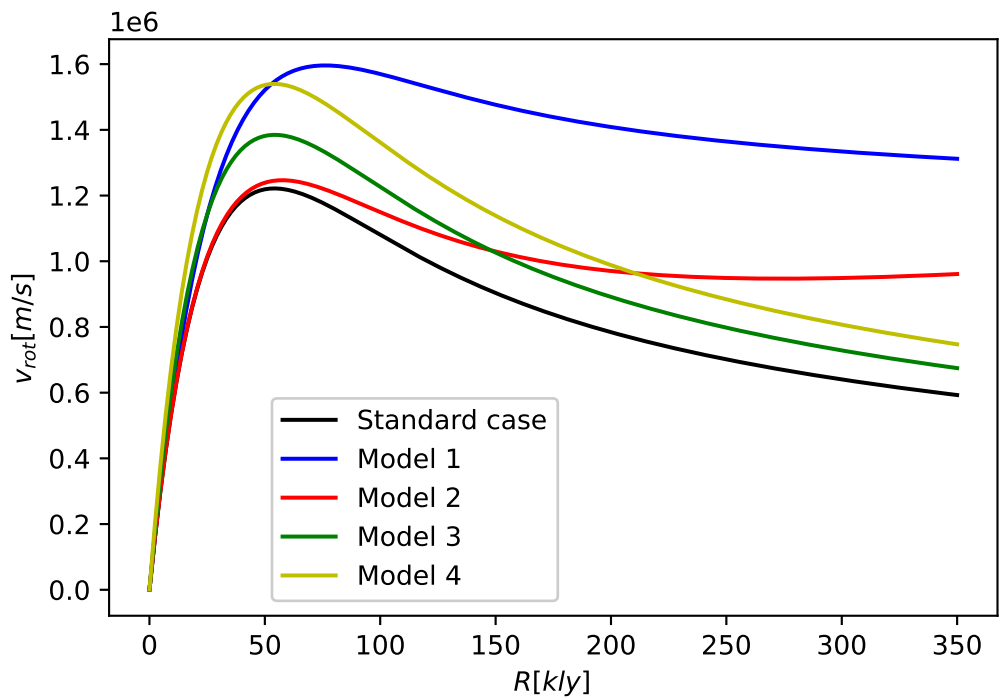}
    \caption{Galactic rotational curves for different models of the $R$ dependence of $\beta_0$.}
    \label{vorbs}
\end{figure}

It turns out that a distance dependence of the GUP parameter $\beta_0$ must be assumed to properly describe the flatness of rotational curves of galaxies.
In fact, as can be noticed from Eq. \eqref{eqn:orbital_velocity_GUP}, a constant GUP parameter simply shifts the orbital velocity at any given radial position by a constant factor.
The assumption that the GUP parameter $\beta_0$ takes a distance dependence is compatible with the fact that similar effects are not observed at the solar system scale, at which Kepler's laws hold, while effects usually associated with DM tend to become relevant approaching galactic and intergalactic scales.
This suggests that the GUP parameter $\beta_0$ must be distance dependent, since DM effects appear to be distance dependent. Such an assumption is also supported by Ref. \cite{Ong:2018zqn} and  the difference between estimations of the quadratic GUP parameter in tabletop experiments, where $\beta_0>0$ \cite{IPK,SLV,KP,Das:2021skl,Bosso:2016ycv}, and astrophysical/cosmological observations, where $\beta_0<0$ \cite{Das:2021nbq,Ong:2018zqn,Nenmeli:2021orl,JizbaScard,BuoninfCorp,JizbaLamb}.
We propose different models of a distance dependent $\beta_0$ 
\begin{itemize}
    \item Model 1: $\displaystyle{\beta_0(R)=\gamma\frac{R}{R^*}}$~,
    \item Model 2: $\displaystyle{\beta_0(R)=\gamma\frac{R^2}{{R^*}^2}}$~,
    \item Model 3: $\displaystyle{\beta_0(R)=\gamma\ln{\left(1+\frac{R}{R^*}\right)}}$~,
    \item Model 4: $\displaystyle{\beta_0(R)=\gamma\frac{2}{\pi}\arctan{\left(\frac{R}{R^*}\right)}}$ 
\end{itemize}
where $\gamma<0$ is a constant parameter and $R^*\sim 1 \mathrm{ ly}$ is the scale at which effects, associated with DM, become significant. We consider a toy model galaxy with the following matter distribution \cite{Freeman:1970mx}
\begin{eqnarray}
\label{matterdist}
\rho(R)=\rho_0e^{-\frac{R}{R_d}}~,
\end{eqnarray}
where we chose $\rho_0=2\times10^{-19}\,\mathrm{kg/m^3}$ for the central density and $R_d=16\,000\,\mathrm{ly}$ for the galaxy scale parameter. We use the matter distribution from Eq. (\ref{matterdist}) to obtain the mass of the galaxy within a certain radius $M(R)$, used in Eq. (\ref{eqn:orbital_velocity_GUP}), to obtain Fig. \ref{vorbs}.

From Fig. \ref{vorbs}, we can see that the model which describes the flatness of rotational curves the best is model 1, since the Compton correction term dominates at large distances, and the linear model 1 renders it constant.
As for the natural logarithm and arc tangent models, they require a much higher peak for the rotational velocities in order to explain the flatness of the curves.
Since observations of rotational curves of galaxies show no significant discrepancy from standard Newton's theory up to about the peak of the curve, such models are not able to fit the observations.
The quadratic model 2 can potentially constitute a good description for a different choice of the parameters $\gamma$ and ${R^*}$.
Since models 3 and 4 do not describe the DM effects satisfactory, we are left with models 1 and 2.
For these models, we can consider the two different parameters $\gamma$ and ${R^*}$ as only one parameter $\gamma/{R^*}$ and $\gamma/{R^*}^2$ for models 1 and 2, respectively.
The values for these parameters, which were used to obtain the above graphs, are $\gamma/{R^*}=-1.9\times10^{33}\,\mathrm{ly^{-1}}$ and $\gamma/{R^*}^2=-3.6\times10^{27}\,\mathrm{ly^{-2}}$ for models 1 and 2, respectively. These values constitute upper bounds for such parameters.
Namely, $|\gamma/{R^*}|\leq1.9\times10^{33}\,\mathrm{ly^{-1}}$ and $|\gamma/{R^*}^2|\leq3.6\times10^{27}\,\mathrm{ly^{-2}}$. 

\section{Conclusion}
\label{conc}

Newton's law of gravity can be derived as an entropic force through the holographic principle \cite{Verlinde:2010hp}.
In the present work, we have revised the derivation considering the influence of GUP.
Specifically, we have considered the influence of GUP on the temperature $T$ in the equipartition theorem, the Bekenstein-Hawking entropy, and the Compton wavelength.
The GUP-corrected law of gravity has then been used to provide an explanation for the flatness of the rotational curves of galaxies.
Specifically, alongside the proposed DM content in galaxies, we proposed that GUP effects can contribute to the observed shape of rotational curves. 
In the case that the GUP parameter $\beta_0$ remains constant, the rotational curves of galaxies only get magnified by a constant factor.
Therefore, for GUP to effectively influence the rotational curves, we argued that the GUP parameter must be distance dependent. 
This claim is directly supported by the work from Ref. \cite{Ong:2018zqn} and indirectly by a comparison of positive bounds of quadratic GUP parameters estimated from laboratory experiments \cite{IPK,SLV,KP,Das:2021skl,Bosso:2016ycv} and negative bounds estimated from astrophysical/cosmological observations  \cite{Das:2021nbq,Ong:2018zqn,Nenmeli:2021orl,JizbaScard,BuoninfCorp,JizbaLamb}.

We have proposed different models concerning the distance dependence for the GUP parameter $\beta_0$ and introduced a new scale parameter ${R^*}$ at which GUP effects start to contribute to the shape of the rotational curves of galaxies.
We note here that the GUP length scale need not be of the order of the Planck length $\ell_p$, but can be any intermediate length scale $\sqrt{\beta_0}\ell_p$ between the electroweak and Planck scales, determined by the GUP parameter $\beta_0$. For the cases of models 1 and 2, we introduced parameters $\gamma/{R^*}$ and $\gamma/{R^*}^2$ respectively, since we cannot obtain bounds for $\gamma$ and ${R^*}$ separately. Models 3 and 4 were found to be inadequate to explain the observed DM effects and there would also be no possibility to combine parameters $\gamma$ and ${R^*}$ in a similar fashion as for models 1 and 2.

Models 1 and 2 constrain the newly defined parameters to $|\gamma/{R^*}|\leq1.9\times10^{33}\,\mathrm{ly^{-1}}$ and $|\gamma/{R^*}^2|\leq3.6\times10^{27}\,\mathrm{ly^{-2}}$, respectively.
However, these models can explain the flatness of rotational curves of galaxies only up to an extent. For example, we notice that for model 1 the rotational velocities would remain constant for $R\to\infty$, while for model 2 they would diverge for $R\to\infty$ for any values of the parameters.
Furthermore, the approximations used to obtain Eqs. (\ref{guptemp}) and (\ref{gravforce}) break down at sufficiently large $R$.

The contribution of GUP effects to the shape of rotational curves of galaxies should be determined once more information on the exact nature of particle DM and its abundance in galaxies is known.
Furthermore, the feature of a distance-dependent GUP parameter, leading towards a partial explanation of galactic rotational curves, can be considered as a possible additional structure of models of quantum mechanics with a minimal length having astrophysical and cosmological consequences.
Finally, it is worth to explore the correspondence between our result and those presented in \cite{Ong:2018zqn,Gnatenko:2019flg,Gnatenko:2019ozg}, which still exhibit the possibility of a mass-dependent GUP parameter.


\end{document}